\documentclass[12pt,cite,epsf,epsfig]{article}
\usepackage{epsfig}

\begin{document}
\author{S. Dev\thanks{dev5703@yahoo.com} , Shivani
Gupta\thanks{shiroberts\_1980@yahoo.co.in}  and  Radha Raman Gautam \thanks{gautamrrg@gmail.com}}
\title{Tribimaximal mixing in neutrino mass matrices with texture zeros or vanishing minors}
\date{\textit{Department of Physics, Himachal Pradesh University, Shimla 171005, India.}\\
\smallskip}
\maketitle
\begin{abstract}
We study the existence of one/two texture zeros or one/two vanishing minors in the neutrino mass matrix with $\mu\tau$ symmetry. In the basis where the charged lepton mass matrix and the Dirac neutrino mass matrix are diagonal, the one/two zeros or one/two vanishing minors in the right-handed Majorana mass matrix having $\mu\tau$ symmetry will propagate via seesaw mechanism as one/two vanishing minors or one/two texture zeros in the neutrino mass matrix with $\mu\tau$ symmetry respectively. It is found that only five such texture structures of the neutrino mass matrix are phenomenologically viable. For tribimaximal mixing, these texture structures reduce the number of free parameters to one. Interesting predictions are obtained for the effective Majorana mass $M_{ee}$, the absolute mass scale and the Majorana-type CP violating phases.
\end{abstract}
\section{Introduction}
It has, now, been firmly established in a variety of solar, atmospheric and terrestrial neutrino oscillation experiments that the neutrinos have masses and they, also, mix like quarks. However, the inferred lepton flavor structure is different from the quark flavor structure. As a result, the Pontecorvo-Maki-Nakagawa-Sakata (PMNS) \cite{1} matrix encoding the lepton mixing and the Cabibbo-Kobayashi-Maskawa (CKM) \cite{2} matrix encoding quark mixing have completely different structures.\\ The lepton mixing can be well described by the so-called tribimaximal mixing matrix \cite{3} which has been taken by many as a hint of an underlying symmetry in the lepton sector. However, some recent global analyses \cite{4} allow for significant deviations from tribimaximal values of mixing angles which disfavors \cite{5} the existence of fundamental tribimaximal (TBM) symmetry. The absence of a plausible model for TBM mixing and the apparent absence of any such symmetry in the quark sector has further aggravated these doubts. To complicate matters further, TBM mixing can be realized in a variety of ways \cite{6} most of which invoke physics at a very high energy scale. In the absence of any generic predictions, it is not possible to test these proposed scenarios most of which do not explain neutrino mass hierarchies. Alternatives to TBM symmetry must, therefore, be explored.
Other alternative scenarios which induce relations between the masses and mixing angles are the existence of texture zeros \cite{7} and zero minors \cite{8} in the neutrino mass matrix. The imposition of the observational constraints of TBM mixing on these, otherwise successfull scenarios leads to some testable predictions which can be used as a check on their phenomenological viability. Thus, combining TBM mixing with texture zeros/zero minors is expected to further our insight into the vexed issue of neutrino masses and mixings. The main motivation for the present study are the relations between masses and mixing angles induced by the existence of texture zeros or zero minors in the neutrino mass matrix. Thus, combining TBM mixing with texture zeros/ zero minors is expected to further our insight into neutrino masses and mixings.
 In the present work, we study the phenomenological implications of the existence of one/two texture zeros or one/two vanishing minors in the neutrino mass matrix with $\mu\tau$ symmetry under the additional observational constraint of TBM mixing and obtain interesting predictions for the effective Majorana mass, absolute neutrino mass scale and the Majorana-type CP violating phases. Some of these textures have been discussed in \cite{9, 10} though in a somewhat different context.\\ 
We reconstruct the neutrino mass matrix in the flavor basis (where the charged lepton mass matrix is diagonal) assuming that the neutrinos are Majorana particles. In this basis the complex symmetric mass matrix $M_{\nu}$ can be diagonalized by unitary matrix V as
\begin{equation}
M_{\nu}=V M_{\nu}^{diag}V^{T}.
\end{equation}
where
\begin{center}
$M_{\nu}^{diag}$ = $\left(
\begin{array}{ccc}
m_{1} & 0 & 0 \\  0& m_{2} & 0 \\ 0& 0 & m_{3}
\end{array}
\right)$
\end{center}
The PMNS mixing matrix $V$ can be decomposed into the unitary matrix U and a phase matrix P.
\begin{equation} 
V=U.P
\end{equation}
where the phase matrix is given as 
\begin{center}
$P = \left(
\begin{array}{ccc}
1 & 0 & 0 \\ 0 & e^{i\alpha} & 0 \\ 0 & 0 & e^{i\beta}
\end{array}
\right)$
\end{center}
 with the two Majorana-type CP violating phases $\alpha$ and $\beta$. Using Eqn. (2), the neutrino mass matrix can be written as
\begin{equation}
M_{\nu}=U P M_{\nu}^{diag}P^{T}U^{T}.
\end{equation}                                                                                             
The unitary matrix U \cite{11} is given by
\begin{equation}
U= \left(
\begin{array}{ccc}
c_{12}c_{13} & s_{12}c_{13} & s_{13}e^{-i\delta} \\
-s_{12}c_{23}-c_{12}s_{23}s_{13}e^{i\delta} &
c_{12}c_{23}-s_{12}s_{23}s_{13}e^{i\delta} & s_{23}c_{13} \\
s_{12}s_{23}-c_{12}c_{23}s_{13}e^{i\delta} &
-c_{12}s_{23}-s_{12}c_{23}s_{13}e^{i\delta} & c_{23}c_{13}
\end{array}
\right)
\end{equation} 
where $s_{ij}=\sin\theta_{ij}$ and $c_{ij}=\cos\theta_{ij}$. 
\section{Formalism}
The present neutrino oscillation data on neutrino masses and mixing angles from solar, atmospheric and reactor neutrino oscillation experiments at 1, 2 and 3$\sigma$ \cite{12} can be summarized as
\begin{eqnarray}
\Delta m_{12}^{2} &=&7.67_{(-0.19,-0.36,-0.53)}^{(+0.16,+0.34,+0.52)}\times
10^{-5}eV^{2}, \nonumber \\ \Delta m_{23}^{2} &=&\pm
2.39_{(-0.8,-0.20,-0.33)}^{(+0.11,+0.27,+0.47)}\times 10^{-3}eV^{2},  \nonumber \\
\theta_{12}& =&33.96_{(-1.12,-2.13,-3.10)}^{o(+1.16,+2.43,+3.80)}, \nonumber \\ \theta_{23}
&=&43.05_{(-3.35,-5.82,-7.93)}^{o(+4.18,+7.83,+10.32)}.  \nonumber
\end{eqnarray}
Moreover, there exists an upper bound on $\theta_{13}$ from the CHOOZ experiment as
\begin{equation}
\theta_{13}< 12.38^{o}(3\sigma).
\end{equation}
Since, our neutrino mass matrix is $\mu\tau$ symmetric, maximal atmospheric neutrino mixing i.e. $s_{23}^2=\frac{1}{2}$ and $s_{13}^2 =0$ immediately follows from this structure of mass matrix. 
Hence, the unitary matrix U which diagonalizes the $\mu\tau$ symmetric mass matrix $M_{\nu}$ takes the form
\begin{equation}
U = \left(
\begin{array}{ccc}
c_{12} & s_{12} & 0 \\
\frac{-s_{12}}{\sqrt{2}}&
\frac{c_{12}}{\sqrt{2}}& \frac{1}{\sqrt{2}}\\
\frac{s_{12}}{\sqrt{2}}&
\frac{-c_{12}}{\sqrt{2}}& \frac{1}{\sqrt{2}}\\
\end{array}
\right)
\end{equation} where $s_{12}=\sin\theta_{12}$ and $c_{12}=\cos\theta_{12}$. The Dirac-type CP violating phase $\delta$ disappears from the mixing matrix U for vanishing $\theta_{13}$. Thus, the only source of CP violation in the leptonic sector are the two physical Majorana-type CP violating phases present in the mixing matrix V. 
The resulting neutrino mass matrix $M_{\nu}$ becomes
\begin{equation}
M_{\nu}= \left(
\begin{array}{ccc}
m_1c_{12}^2+m_2s_{12}^2e^{2i\alpha} & \frac{(-m_1+m_2 e^{2i\alpha})c_{12}s_{12}}{\sqrt{2}} & \frac{(-m_1+m_2 e^{2i\alpha})c_{12}s_{12}}{\sqrt{2}} \\
\frac{(-m_1+m_2 e^{2i\alpha})c_{12}s_{12}}{\sqrt{2}} &
\frac{1}{2}\left( m_1s_{12}^2+m_2c_{12}^2e^{2i\alpha}+m_3e^{2i\beta}\right) & \frac{1}{2}\left( m_1s_{12}^2+m_2c_{12}^2e^{2i\alpha}-m_3e^{2i\beta}\right) \\
\frac{(-m_1+m_2 e^{2i\alpha})c_{12}s_{12}}{\sqrt{2}}&
\frac{1}{2}\left( m_1s_{12}^2+m_2c_{12}^2e^{2i\alpha}-m_3e^{2i\beta}\right) & \frac{1}{2}\left( m_1s_{12}^2+m_2c_{12}^2e^{2i\alpha}+m_3e^{2i\beta}\right)
\end{array}
\right)
\end{equation} 
This matrix is parameterized in terms of three
neutrino mass eigenvalues ($m_1, m_2, m_3$), one neutrino mixing
angle ($\theta _{12}$) i.e. solar mixing angle 
and the two Majorana-type CP violating phases $\alpha$ and $\beta$.
The masses $m_2$ and $m_3$ can be calculated from the mass-squared differences $\Delta m_{12}^2$ and $\Delta m_{23}^2$ using the relations
\begin{equation}
m_{2}=\sqrt{m_{1}^{2}+\Delta m_{12}^{2}}
\end{equation}
and
\begin{equation}
m_{3}=\sqrt{m_{2}^{2}+\Delta m_{23}^{2}}.
\end{equation}
Additional imposition of tribimaximal mixing restricts the solar mixing angle i.e. $s_{12}^2=\frac{1}{3}$. 
All available information on neutrino masses from solar, atmospheric neutrino experiments is encoded in the parameter 
\begin{equation}
R_{\nu}=\frac{m_{2}^{2}-m_{1}^{2}}{|m_{3}^{2}-m_{2}^{2}|}
\end{equation}
In addition, cosmological observations put an upper bound on the parameter 
\begin{equation}
\Sigma=\sum_{i=1}^{3}m_i
\end{equation}
which is the sum of the neutrino masses. WMAP data \cite{13} limit $\Sigma$ to be less than (0.6)eV at 95 \% C.L.
The observation of  neutrinoless double beta decay would imply lepton number violation and Majorana nature for neutrinos. The effective Majorana mass of the electron neutrino $M_{ee}$ which
 determines the rate of neutrinoless double beta decay \cite{14} is given by
\begin{equation}
M_{ee}= |m_1c_{12}^2c_{13}^2+ m_2s_{12}^2c_{13}^2e^{2i\alpha}+m_3s_{13}^2e^{2i(\beta - \delta)}|
\end{equation}
which in the limit of vanishing $\theta_{13}$ becomes
\begin{equation}
M_{ee}= |m_1c_{12}^2+ m_2s_{12}^2e^{2i\alpha}|.
\end{equation}
The possible measurement of the effective Majorana mass in the neutrinoless double beta decay experiments will provide an additional constraint on the neutrino mass scale and the two Majorana-type CP violating phases. Thus, the analysis of $M_{ee}$ will be significant. A stringent constraint $|M_{ee}|< 0.35$eV was obtained by the $^{76}Ge$
 Heidelberg-Moscow experiment \cite{15}.
 There are large number of projects such as SuperNEMO\cite{15},
 CUORE\cite{16}, CUORICINO\cite{16} and  GERDA\cite{17} which
 aim to achieve a sensitivity below 0.01eV on $M_{ee}$.
 Forthcoming experiment SuperNEMO, in particular, will explore
 $M_{ee}$ $<$ 0.05eV\cite{18} 
The neutrino mass scale will be independently determined by the direct beta decay searches \cite{19} and cosmological observations. 
We, therefore, present some interesting predictions on the two parameters $M_{ee}$ and masses of neutrinos which can, hopefully, be tested in the near future. 
The neutrino mass matrix $M_{\nu}$ in Eqn. (7) is clearly $\mu\tau$ symmetric having two physical phases $2\alpha$ and $2\beta$. $\mu\tau$ symmetry constrains the two mixing angles $\theta_{13}$ and $\theta_{23}$ but imposes no constraints whatsoever on the neutrino masses and the two Majorana phases. On the other hand, textures induce linear relations between mass matrix elements, thereby, relating the mixing matrix elements to the neutrino masses and the Majorana phases. Imposition of $\mu\tau$ symmetry on textures will, therefore, lead to highly predictive scenarios for the unknown parameters. In the present work, we study the phenomenological implications of a class of neutrino mass matrices with $\mu\tau$ symmetry having one/two texture zeros or one/two zero minors under the condition of tribimaximality which reduces the number of free parameters to just one which can be identified with the neutrino mass scale $m_1$. We test the phenomenological viability of this class of texture structures by randomly generating $m_1$ whose upper bound is taken from the WMAP data while the mass-squared differences are varied over their $3\sigma$ experimental ranges. 
\section{Results and Discussions}
The seesaw mechanism for understanding the scale of neutrino masses is
regarded as the prime candidate not only due to its simplicity but
also due to its theoretical appeal. In the framework of type-I
seesaw mechanism \cite{20}, the effective Majorana mass matrix
$M_{\nu}$ is given by
\begin{equation}
 M_{\nu}= - M_D M_R^{-1} M_D^T  \nonumber \\
\end{equation}
where $M_D$ is the Dirac neutrino mass matrix and $M_R$ is the
right-handed Majorana mass matrix. $M_{\nu}$ will be $\mu\tau$  symmetric if both $M_D$ and $M_R$ are $\mu\tau$ symmetric. 

\subsection{Neutrino mass matrices with one/two vanishing minors and $\mu\tau$ symmetry}
In the basis where the charged lepton mass matrix and the Dirac-type neutrino mass matrix are diagonal, the Dirac-type neutrino mass matrix under $\mu\tau$ symmetry takes the form $M_D=diag(a,b,b)$. The assumption of diagonal Dirac neutrino mass matrices is well motivated from symmetry considerations and there exist models of neutrino masses/mixing based on discrete \cite{21, 22} as well as continuous \cite{23} symmetries which naturally lead to diagonal Dirac mass matrices for neutrinos. The one/two zeros in the Majorana mass matrix $M_R$ having $\mu\tau$ symmetry will show as the one/two vanishing minors of $M_{\nu}$ with $\mu\tau$ symmetry. Since it is easy to visualize $\mu\tau$ symmetry in a mass matrix having texture zeros than vanishing minors, we find all possible texture structures of $M_R$ with one/two zeros and $\mu\tau$ symmetry. Out of total twenty one possible  one/two zeros textures in $M_R$ only five are found to be compatible with $\mu\tau$ symmetry. All five possible structures of $M_R$ and corresponding $M_{\nu}$  are given in Table 1.
\begin{table}[h]
\begin{center}
\begin{small}
\begin{tabular}{|c|c|c|}
\hline    & $M_R$ & $M_{\nu}$ \\
\hline 1. & $\left(
\begin{array}{ccc}
0 & B & B \\  B & C & D \\ B& D & C
\end{array}
\right)$ & $\left(
\begin{array}{ccc}
\frac{-a^2(C+D)}{2B^2} & \frac{ab}{2B}& \frac{ab}{2B}\\  \frac{ab}{2B} & \frac{b^2}{2C-2D} & \frac{b^2}{-2C+2D} \\\frac{ab}{2B}& \frac{b^2}{-2C+2D} & \frac{b^2}{2C-2D}
\end{array}
\right)$   \\
\hline 2. &$\left(
\begin{array}{ccc}
A & B & B \\  B& C & 0 \\ B& 0 & C
\end{array}
\right)$   &$\left(
\begin{array}{ccc}
\frac{a^2C}{-2B^2+AC} & \frac{abB}{2B^2-AC} &  \frac{abB}{2B^2-AC} \\   \frac{abB}{2B^2-AC} & \frac{b^2(B^2-AC)}{2B^2C-AC^2} & \frac{b^2B^2}{-2B^2C+AC^2} \\ \frac{abB}{2B^2-AC}& \frac{b^2B^2}{-2B^2C+AC^2} & \frac{b^2(B^2-AC)}{2B^2C-AC^2}
\end{array}
\right)$  \\
\hline 3. & $\left(
\begin{array}{ccc}
A & B & B \\ B & 0 & D \\B & D  & 0
\end{array}
\right)$ & $\left(
\begin{array}{ccc}
\frac{a^2D}{-2B^2+AD} & \frac{abB}{2B^2-AD} & \frac{abB}{2B^2-AD} \\  \frac{abB}{2B^2-AD} & \frac{b^2B^2}{-2B^2D+AD^2} & \frac{b^2(B^2-AD)}{2B^2D-AD^2}\\ \frac{abB}{2B^2-AD}&\frac{b^2(B^2-AD)}{2B^2D-AD^2} &\frac{b^2B^2}{-2B^2D+AD^2}
\end{array}
\right)$   \\
\hline 4. & $\left(
\begin{array}{ccc}
A & 0 & 0 \\ 0 & C & D \\0 & D & C
\end{array}
\right)$ & $\left(
\begin{array}{ccc}
\frac{a^2}{A} & 0 & 0 \\  0 & \frac{b^2C}{C^2D^2} & \frac{b^2D}{-C^2+D^2} \\ 0& \frac{b^2D}{-C^2+D^2}& \frac{b^2C}{C^2D^2}
\end{array}
\right)$   \\
\hline 5. &$\left(
\begin{array}{ccc}
0 & B & B \\  B & C & 0 \\ B & 0 & C
\end{array}
\right)$  & $\left(
\begin{array}{ccc}
\frac{-a^2C}{2B^2} & \frac{ab}{2B} &  \frac{ab}{2B}\\  \frac{ab}{2B} &\frac{b^2}{2C}&\frac{-b^2}{2C} \\  \frac{ab}{2B}& \frac{-b^2}{2C} & \frac{b^2}{2C}
\end{array}
\right)$  \\
\hline
\end{tabular}
\end{small}
\caption{Texture structures of $M_R$ with one/two texture zeros and $M_\nu$ with one/two vanishing minors and $\mu\tau$ symmetry.}
\end{center}
\end{table}

These texture structures of $M_R$ with a diagonal $M_D$ result in $M_{\nu}$ with one/two vanishing minors in addition to $\mu\tau$ symmetry. However, texture 4 and 5 of $M_R$ imply neutrino mass matrices which are phenomenologically not viable. Therefore, we study the phenomenological implications of the remaining three viable texture structures of the neutrino mass matrix $M_{\nu}$ and obtain interesting implications. \\
\begin{large}\textbf{Case 1}\end{large}\\
Texture 1 of $M_R$ with diagonal $M_D$ leads via type-I seesaw to neutrino mass matrix $M_{\nu}$ where the minor corresponding to (1,1) element vanishes.
\begin{equation}
M_{\nu(22)}M_{\nu(33)}-M_{\nu(23)}M_{\nu(32)}=0.  
\end{equation} \nonumber
Substituting the values of mass matrix elements  from Eqn. (7), Eqn. (15) takes the form 
\begin{equation}
m_1s_{12}^2+m_2c_{12}^2e^{2i\alpha}=0.
\end{equation}
which gives
\begin{equation}
\left|\frac{m_1}{m_2}\right|=\frac{1}{tan^2\theta_{12}}.
\end{equation}
Hence, this texture of $M_{\nu}$ with a vanishing minor and  $\mu\tau$ symmetry is disallowed by the solar mass hierarchy.\\
\begin{large}\textbf{Case 2}\end{large}\\
Texture 2 of $M_R$ results in $M_{\nu}$ where the minor corresponding to (2,3) element vanishes. The realization of such textures of $M_R$ can be done by discrete non-Abelian symmetry $D_4$, studied by Grimus and Lavoura \cite{21}. The equation of vanishing minor in this case becomes
\begin{equation}
M_{\nu(11)}M_{\nu(32)}-M_{\nu(12)}M_{\nu(31)}=0.  
\end{equation}
Using the values of the elements of $M_{\nu}$ from Eqn. (7) leads to
\begin{equation}
-\frac{1}{m_3}e^{-2i\beta}+\frac{1}{m_2}c_{12}^2e^{-2i\alpha}+\frac{1}{m_1}s_{12}^2=0. 
\end{equation}
Under the condition of tribimaximal mixing i.e. $s_{12}^2=\frac{1}{3}$, the above equation becomes
\begin{equation}
-\frac{1}{m_3}e^{-2i\beta}+\frac{2}{3m_2}e^{-2i\alpha}+\frac{1}{3m_1}=0. 
\end{equation}
Separating Eqn. (20) into real and imaginary parts we get
\begin{eqnarray}
-\frac{1}{m_3}cos2\beta+\frac{2}{3 m_2}cos2\alpha+\frac{1}{3m_1}=0 \nonumber, \\
-\frac{1}{m_3}sin2\beta+\frac{2}{3 m_2}sin2\alpha =0.
\end{eqnarray}
We note that the two Majorana phases $2\alpha$ and $2\beta$ are correlated to each other. Hence, there is only one independent phase. Choosing this free phase to be $2\alpha$, we obtain
\begin{equation}
cos2\alpha=\frac{\left(\frac{3m_1m_2}{m_3}\right)^2-4m_1^2-m_2^2}{4m_1m_2}
\end{equation}
The value of $cos2\alpha$ becomes greater than (less than) 1 (-1) for $m_3 < m_1$. Therefore, this texture implies a normal hierarchical mass spectrum. 
It can be seen that this phase becomes the function of a single parameter $m_1$. This phase appears in the expression for $M_{ee}$ which determines the rate of neutrinoless double beta decay.  
The allowed range of effective Majorana mass in this case is $(7.0\times 10^{-4} <M_{ee}<0.2)eV$ and a bound on the lowest mass $m_1>(0.003)eV$ at 3$\sigma$ is obtained.
Figure 1(a) shows the correlation of the lowest mass $m_1$ and effective Majorana mass $M_{ee}$. The normal hierarchy of this texture can be seen from figure 1(b). The Majorana phase $2\alpha$ is found to be constrained in the range ($20^o-180^o$). Hence, there is only one free parameter $m_1$ left while the other parameters $m_2$, $m_3$ and the phase $2\alpha$ can be determined  from $m_1$. 
\begin{figure}
\begin{center}
\epsfig{file=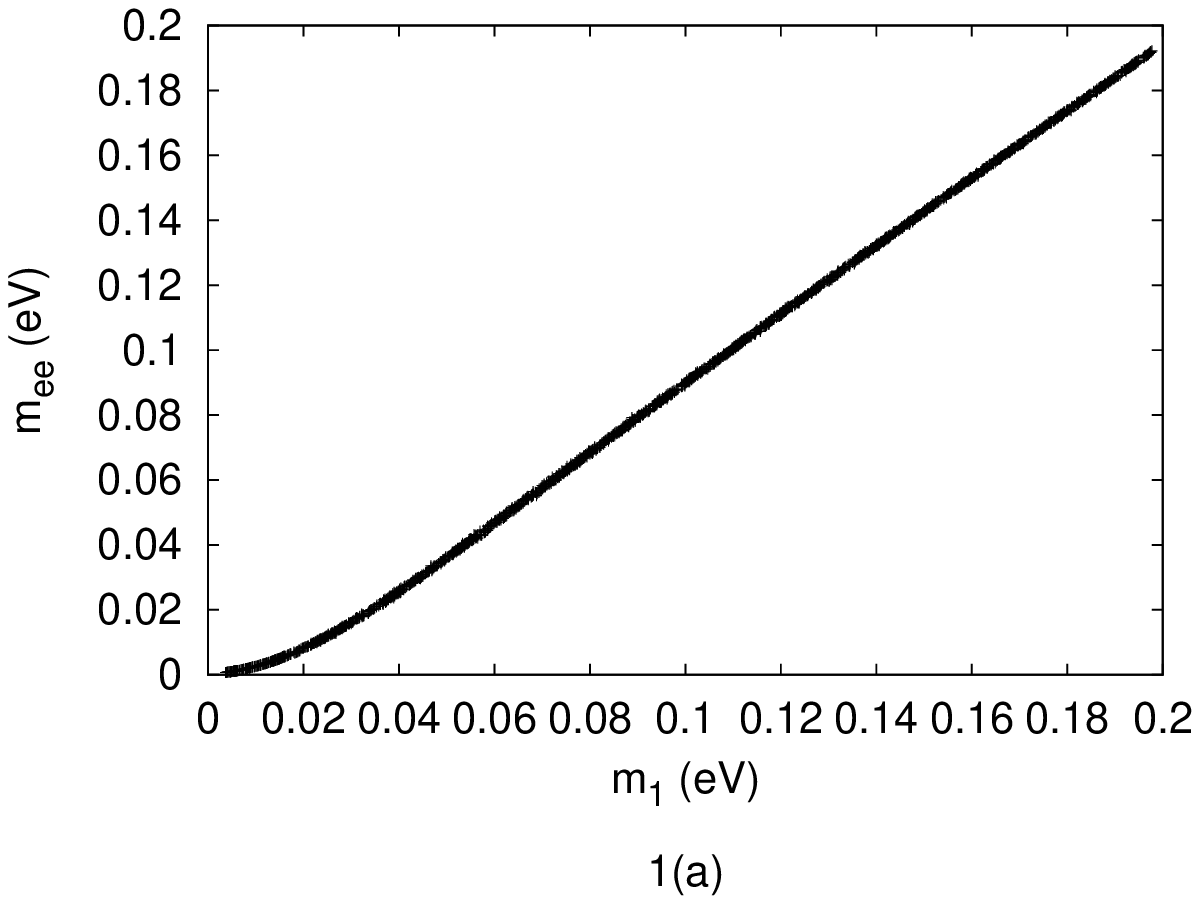,height=5.0cm,width=5.0cm}
\epsfig{file=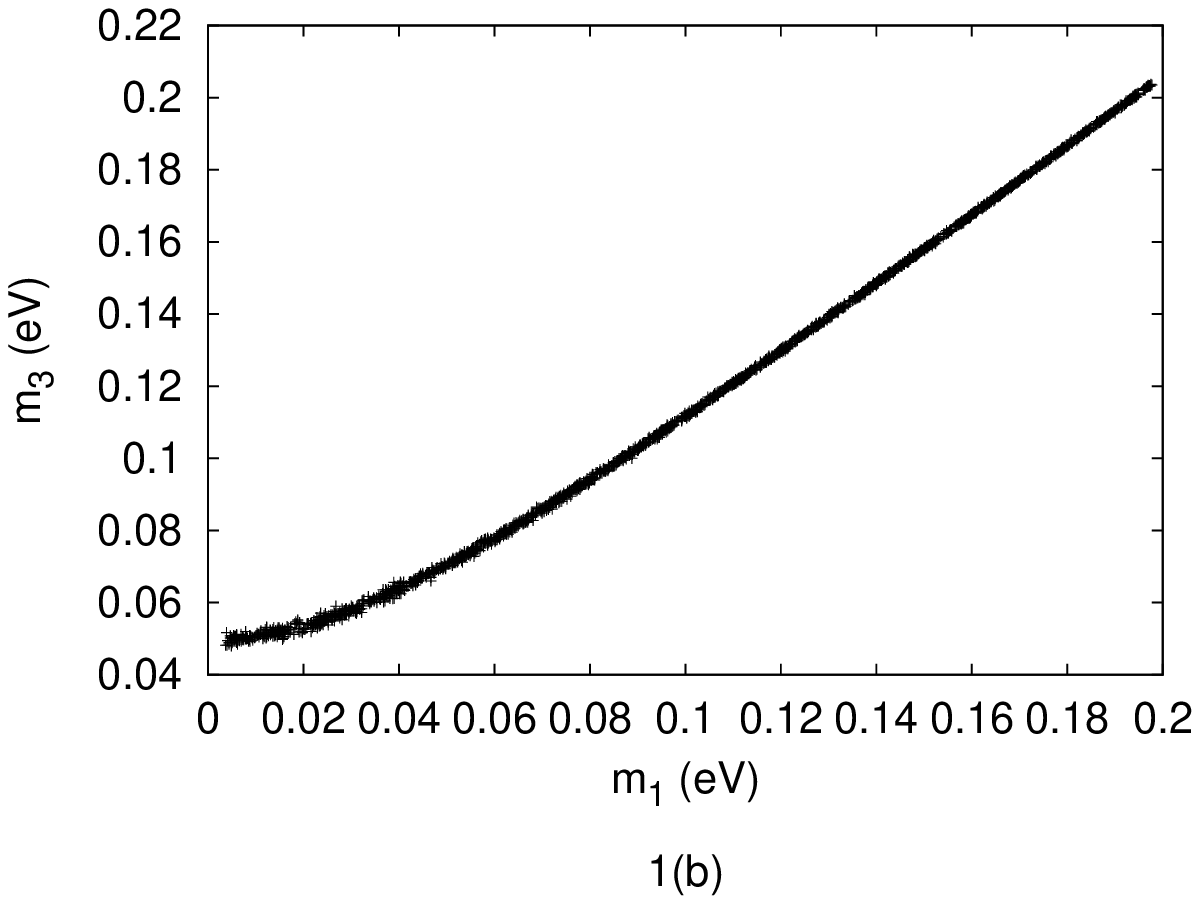,height=5.0cm,width=5.0cm}
\end{center}
\caption{Correlation plots for texture structures with one/two vanishing minors and tribimaximal mixing.}
\end{figure}
\\
\begin{large}\textbf{Case 3}\end{large}\\
Texture 3 of $M_R$ results in $M_{\nu}$ with $\mu\tau$ symmetry and having vanishing minors corresponding to (2,2) and (3,3) elements. This structure of $M_R$ can be realized from discrete non-Abelian $S_3$ $\times$ $Z_2$ \cite{22} symmetry. 
The equations of vanishing minors in this case become
\begin{eqnarray} 
M_{\nu(11)}M_{\nu(33)}-M_{\nu(13)}M_{\nu(31)}=0,  \nonumber\\
M_{\nu(11)}M_{\nu(22)}-M_{\nu(12)}M_{\nu(21)}=0.  
\end{eqnarray}
which leads to 
\begin{equation}
\frac{1}{m_3}e^{-2i\beta}+\frac{1}{m_2}c_{12}^2e^{-2i\alpha}+\frac{1}{m_1}s_{12}^2=0. 
\end{equation}
Under tribimaximal mixing i.e. $s_{12}^2=\frac{1}{3}$, the above equation becomes
\begin{equation}
\frac{1}{m_3}e^{-2i\beta}+\frac{2}{3m_2}e^{-2i\alpha}+\frac{1}{3m_1}=0. 
\end{equation}
Texture 3 of $M_{\nu}$ has identical predictions for the mass eigenvalues and the Majorana-type CP violating phase $2\alpha$ as texture 2. Hence, these two texture structures are equivalent \cite{22}.
\subsection{Neutrino mass matrices with one/two texture zeros and $\mu\tau$ symmetry.}
$\mu\tau$ symmetry can, also, be imposed on the texture structures of $M_R$ having vanishing one/two minors. In the charged lepton basis, where Dirac-type neutrino mass matrix is diagonal  $M_D=diag(a,b,b)$, imposing $\mu\tau$ symmetry on $M_R$ with one/two vanishing minors will show as the one/two texture zeros in the neutrino mass matrix $M_{\nu}$. Out of total twenty one possible  one/two minors in $M_R$, only three are found to be compatible with $\mu\tau$ symmetry. The phenomenologically viable texture structures of $M_R$ and the corresponding $M_{\nu}$  are given in Table 2.
\begin{table}[h]
\begin{center}
\begin{small}
\begin{tabular}{|c|c|c|}
\hline    & $M_R$ & $M_{\nu}$ \\
\hline 1. & $\left(
\begin{array}{ccc}
A & B & B \\  B & C & -C \\ B& -C & C
\end{array}
\right)$ & $\left(
\begin{array}{ccc}
0 & \frac{ab}{2B}& \frac{ab}{2B}\\  \frac{ab}{2B} & \frac{b^2(B^2-AC)}{4B^2C} & \frac{-b^2(B^2+AC)}{4B^2C} \\\frac{ab}{2B}&\frac{-b^2(B^2+AC)}{4B^2C} & \frac{b^2(B^2-AC)}{4B^2C}
\end{array}
\right)$   \\
\hline 2. &$\left(
\begin{array}{ccc}
\frac{B^2}{D} & B & B \\  B& C & D \\ B& D & C
\end{array}
\right)$   & $\left(
\begin{array}{ccc}
\frac{a^2D(C+D)}{B^2(C-D)} & \frac{abD}{B(D-C)} &  \frac{abD}{B(D-C)} \\   \frac{abD}{B(D-C)} & \frac{b^2}{(C-D)} & 0 \\\frac{abD}{B(D-C)}& 0 & \frac{b^2}{(C-D)}
\end{array}
\right)$  \\
\hline 3. & $\left(
\begin{array}{ccc}
\frac{B^2}{C}& B & B \\ B & C & D \\B & D  & C
\end{array}
\right)$ & $\left(
\begin{array}{ccc}
\frac{-a^2C(C+D)}{B^2(C-D)} & \frac{abC}{B(C-D)} & \frac{abC}{B(C-D)} \\  \frac{abC}{B(C-D)} & 0 & \frac{b^2}{D-C}\\ \frac{abC}{B(C-D)}&\frac{b^2}{D-C} &0
\end{array}
\right)$   \\
\hline
\end{tabular}
\end{small}
\caption{Texture structures of $M_R$ with one/two vanishing minors and $M_\nu$ with one/two texture zeros and $\mu\tau$ symmetry.}
\end{center}
\end{table}  
We study the phenomenological implications of these three viable texture structures and obtain interesting constraints on the unknown parameters in the neutrino mass matrix $M_{\nu}$.\\ 
\begin{large}\textbf{Case 1}\end{large}\\
Texture 1 of $M_{\nu}$ has a texture zero at (1,1) place in addition to $\mu\tau$ symmetry. Texture zero condition results in the complex equation
\begin{equation}
m_1c_{12}^2 + m_2 s_{12}^2e^{2i\alpha}=0.
\end{equation}
which vanishes if
\begin{equation}
2\alpha=\left( n+\frac{1}{2}\right)\pi
\end{equation}
and
\begin{equation}
\left| \frac{m_1}{m_2}\right|=tan^2\theta_{12}.
\end{equation}
For tribimaximal mixing, the mass ratio becomes
\begin{equation}
\left| \frac{m_1}{m_2}\right|=\frac{1}{2}.
\end{equation}
The mass eigenvalue $m_1$ is given by
\begin{equation}
m_1=\sqrt{\frac{\Delta m_{12}^2}{3}}
\end{equation}
both for normal and inverted hierarchies. However, as shown in \cite{24}, it turns out that interesting implications from texture zero at $M_{ee}$ only occur for normal hierarchical mass spectrum. 
The only free parameter $m_1$ can be determined within the errors of mass-squared difference $\Delta m^2_{12}$ while the CP violating phase $2\alpha$ is fixed.\\ 
\begin{large}\textbf{Case 2}\end{large}\\
Texture 2 of $M_{\nu}$ in Table 2 has a texture zero at (2,3) place in addition to $\mu\tau$ symmetry. The condition of texture zero implies
\begin{equation}
m_3e^{2i\beta}-m_2c_{12}^2e^{2i\alpha}-m_1s_{12}^2=0.
\end{equation}
For tribimaximal mixing the above equation becomes
\begin{equation}
m_3e^{2i\beta}-\frac{2}{3}m_2e^{2i\alpha}-\frac{1}{3}m_1=0.
\end{equation}
 Separating Eqn. (32) into real and imaginary parts we obtain
\begin{eqnarray}
-m_3cos2\beta + \frac{2}{3}m_2cos2\alpha+ \frac{1}{3}m_1=0, \nonumber \\
-m_3sin2\beta +\frac{2}{3} m_2sin2\alpha =0. 
\end{eqnarray}

It can be seen that the two CP violating phases $2\alpha$ and $2\beta$ are correlated to each other. Taking $2\alpha$ to be the independent phase, we can simultaneously, solve the above two equations to obtain
\begin{equation}
cos2\alpha=\frac{(3m_3)^2-(2m_2)^2-m_1^2}{4m_1m_2}.
\end{equation}
The value of $cos2\alpha$ becomes greater than (less than) 1 (-1) for $m_1 < m_3$. As all the parameters in the above equation can be determined in terms of a single parameter $m_1$, hence, the CP violating phase $2\alpha$ is a function of single parameter $m_1$.
Tribimaximal mixing constrains the lowest mass $m_1$ to the range $(0.05<m_1<0.2)eV$ and the effective Majorana mass to a range $(0.01<M_{ee}<0.2)eV$ at 3$\sigma$ (Fig. 2(a)). The inverted  mass hierarchy for this texture structure can be seen from Figure 2(b). The phase $2\alpha$ is constrained in the range $(20^o-180^o)$ and a strong correlation with the other Majorana-type phase $2\beta$ can be seen from Figure 2(c). \\ 
\begin{figure}
\begin{center}
\epsfig{file=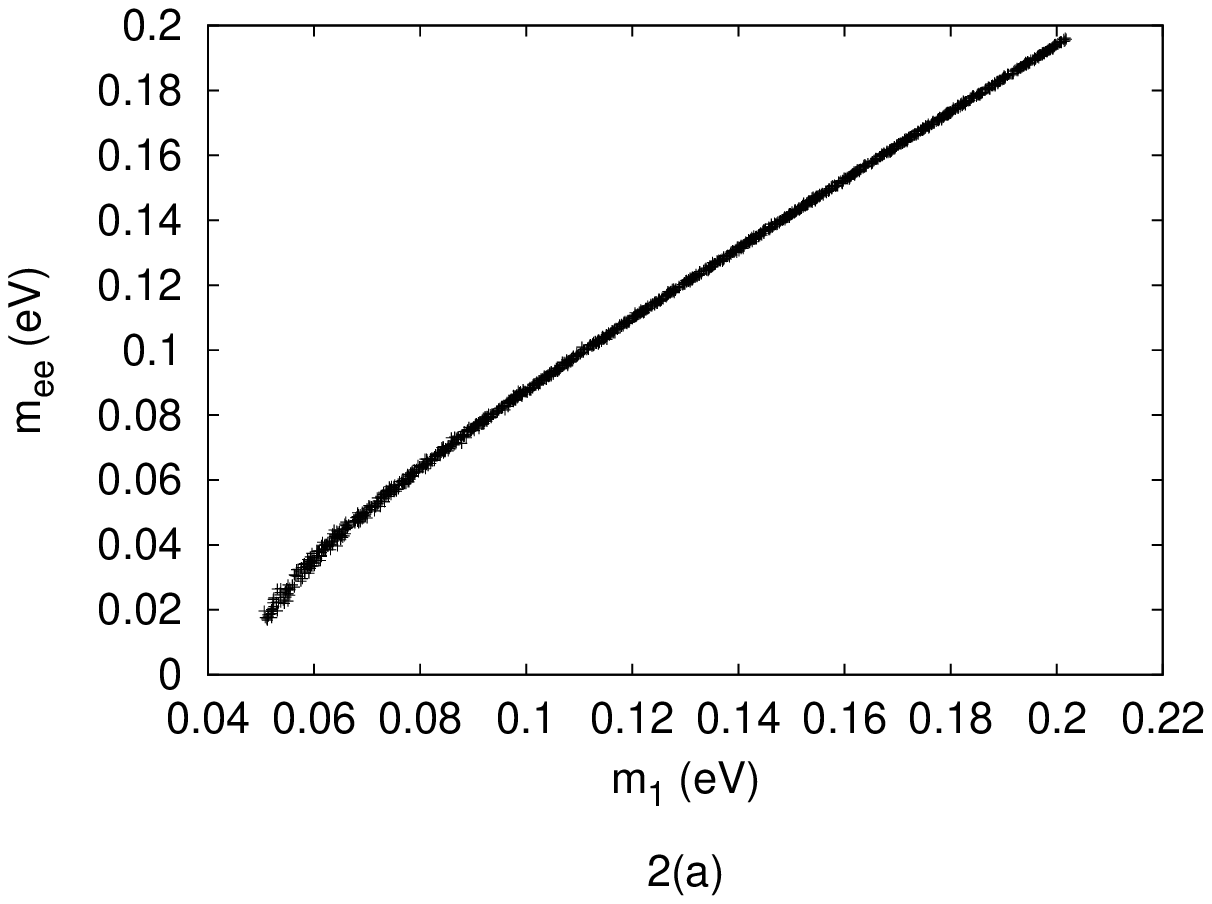,height=5.0cm,width=5.0cm}
\epsfig{file=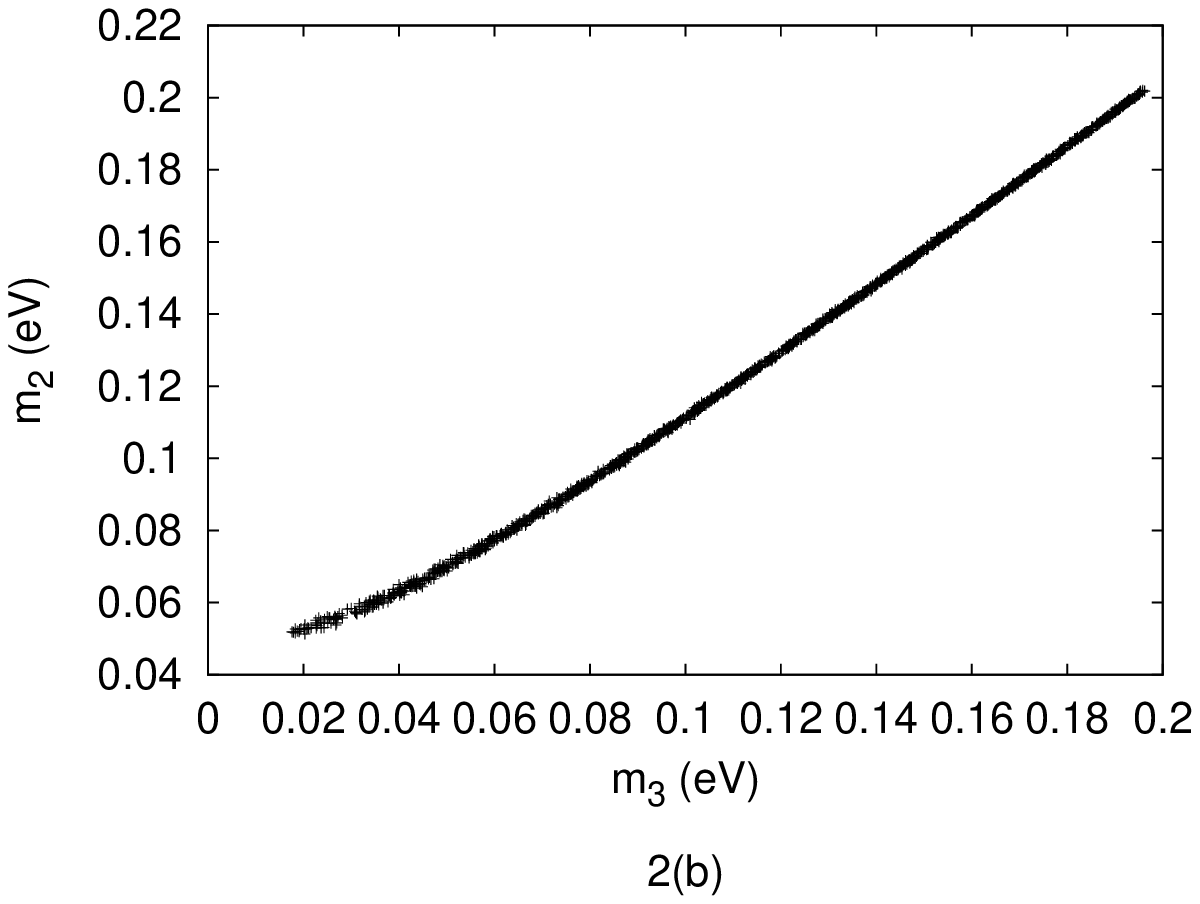,height=5.0cm,width=5.0cm} 
\epsfig{file=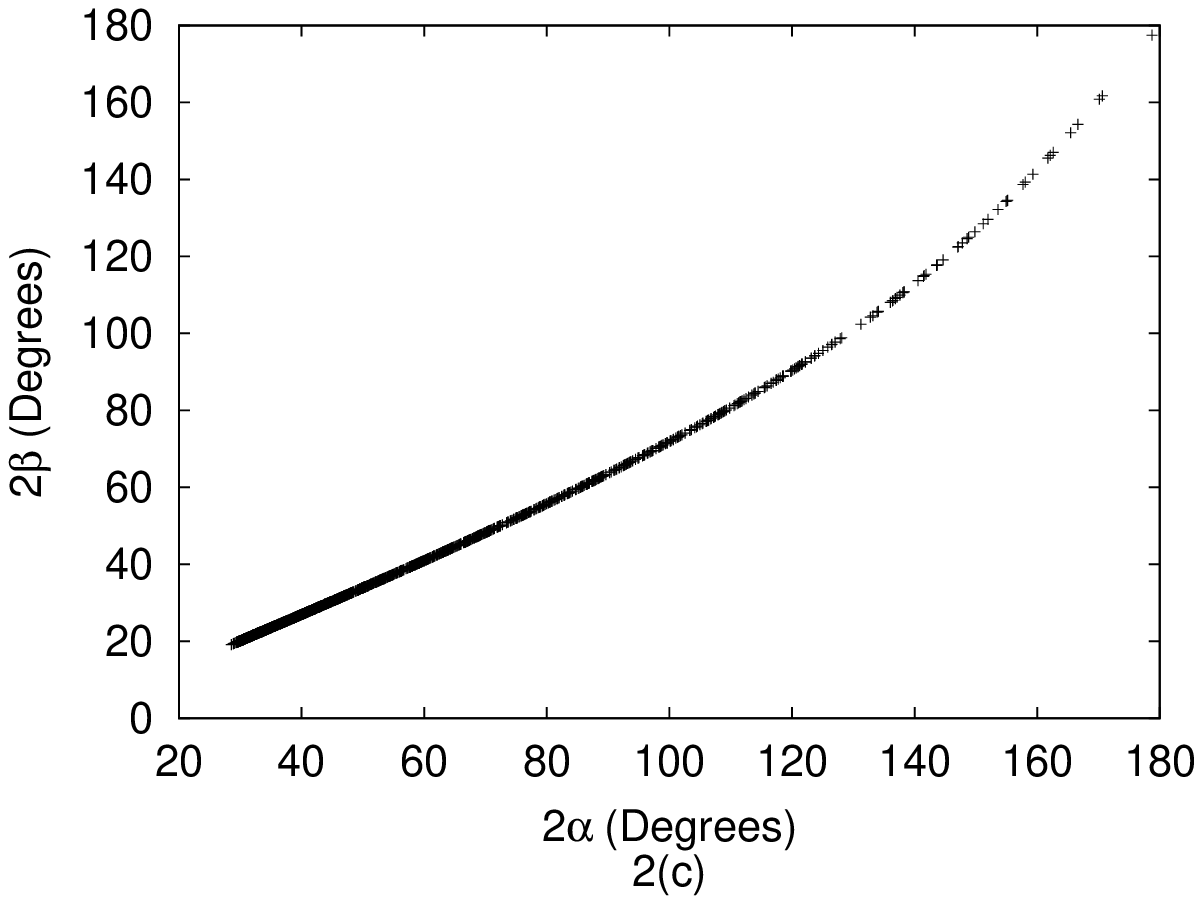,height=5.0cm,width=5.0cm} 
\caption{Correlation plots for texture structures with one/two texture zeros and tribimaximal mixing.} 
\end{center}
\end{figure}\\
\begin{large}\textbf{Case 3}\end{large}\\
Texture 3 of $M_{\nu}$ in Table 2 has two texture zeros at (2,2) and (3,3)
places which lead to the complex equation
\begin{equation}
m_1s_{12}^2 + m_2c_{12}^2e^{2i\alpha}+m_3e^{2i\beta}=0.
\end{equation}
This class of textures has been extensively studied \cite{25} in the literature. However, for tribimaximal mixing this texture structure has the same phenomenological implications as texture 2 and, thus, both are equivalent.

Since in seesaw models any symmetry of the neutrino mass matrix is supposed to manifest only at the seesaw scale, it is of interest to know how the radiative corrections will affect these predictions. It is, however, well known that the radiative corrections are small in case of normal or inverted hierarchical mass spectrum. Since all the texture structures investigated here for tribimaximal mixing give either normal or inverted mass hierarchical spectrum, the radiative corrections will be rather small \cite{26}. Hence, the predictions for the parameters of the neutrino mass matrix are expected to be quite stable.
\section{Conclusions} 
We presented a comprehensive analysis of different texture structures of the neutrino mass matrix that can be obtained via type-I seesaw mechanism from heavy right-handed Majorana mass matrix $M_R$ having one/two vanishing minors or one/two texture zeros. When $\mu\tau$ symmetry is imposed on such $M_R$, it results in $\mu\tau$ symmetric $M_{\nu}$ having one/two texture zeros or one/two vanishing minors via seesaw mechanism. It is found that there are only five possible texture structures of $M_R$ with one/two vanishing minors or one/two texture zeros and are phenomenologically viable. We study the phenomenological implications of all the five texture structures and obtain interesting results for the unknown parameters such as the effective neutrino mass and the neutrino masses. All five textures are found to have a hierarchical mass spectrum. Imposition of tribimaximality reduces the number of free parameters to one. Only one of the two Majorana-type CP violating phases appears in the effective Majorana mass $M_{ee}$ which determines the rate of neutrinoless double beta decay. Predictions for the parameters $M_{ee}$ and absolute mass scale are given for all allowed texture structures. These parameters are expected to be measured in the forthcoming neutrino oscillation experiments. Due to the hierarchical mass spectrum of these textures, the radiative corrections are expected to be negligibly small.

\textbf{\textit{\Large{Acknowledgements}}} \\
The research work of S. D. is supported by the University Grants
Commission, Government of India \textit{vide} Grant No. 34-32/2008
(SR).  S. G. and R. R. G. acknowledge the financial support provided by the Council for Scientific and Industrial Research (CSIR), Government of India.


\begin{thebibliography}{99}
\bibitem{1} B. Pontecorvo,  \textit{Zh. Eksp. Teor. Fiz. (JTEP)} \textbf{33}, 549 (1957); \textit{ibid.} \textbf{34}, 247 (1958);  \textit{ibid.} \textbf{53}, 1717 (1967); Z. Maki, M. Nakagawa and S. Sakata, \textit{Prog. Theor. Phys.} \textbf{28}, 870 (1962).
\bibitem{2} N. Cabibbo,  \textit{Phys. Rev. Lett.} \textbf{10}, 531 (1963); M. Kobayashi and T. Maskawa, \textit{Prog. Theor. Phys.} \textbf{49}, 652 (1973). 
\bibitem{3} P. F. Harrison, D. H. Perkins and W. G. Scott, \textit{Phys. Lett.} \textbf{B 530}, 167 (2002), hep/ph 0202074; P. F. Harrison and W. G. Scott, \textit{Phys. Lett.} \textbf{B 535}, 163 (2002), hep/ph 0203209.
\bibitem{4} G. L. Fogli, E. Lisi, A. Marrone, A. Palazzo and A. M. Rottunno, \textit{Phys. Rev. Lett.} \textbf{101}, 141801 (2008), hep-ph/0806.2649; M. C. Gonzalez-Garcia, M. Maltoni and J. Salvado, hep-ph/1001.4524, R. Wandell \textit{et al.}, arXiv: hep-ex/1002.3471
\bibitem{5} Mohammed Abbas, A. Yu. Smirnov, hep-ph/ 1004.0099; Carl H. Albright, Alexander Dueck and Werner Rodejohann, hep-ph/ 1004.2798 
\bibitem{6} A. Y. Smirnov, hep-ph/0402264; M. Raidal, \textit{Phys. Rev. Lett.} \textbf{93}, 161801 (2004) hep-ph/0404046; H. Minakata and A. Y. Smirnov, \textit{Phys. Rev.} \textbf{D 70}, 073009 (2004), hep-ph/0405088; L. Merlo, \textit{Acta. Phys. Polon.} \textbf{B}, 3179 (2009), hep-ph/0910.2810
\bibitem{7} Paul H. Frampton, Sheldon L. Glashow and Danny
Marfatia, \textit{Phys. Lett.} \textbf{B 536}, 79 (2002), hep-ph/0201008; Bipin R. Desai, D. P. Roy and Alexander R. Vaucher, \textit{Mod. Phys.
Lett.} \textbf{A 18}, 1355 (2003), hep-ph/0209035; Wanlei Guo and Zhi-zhong Xing, \textit{Phys. Rev.} \textbf{D 67},
053002 (2003), hep-ph/0212142; S. Dev, Sanjeev Kumar, Surender Verma and Shivani Gupta,
 \textit{Nucl. Phys.} \textbf{B 784}, 103-117 (2007), hep-ph/0611313; S. Dev, Sanjeev Kumar, Surender Verma and Shivani Gupta,
 \textit{Phys. Rev.} \textbf{D 76}, 013002 (2007), hep-ph/0612102; G. Ahuja, S. Kumar, M. Randhawa, M. Gupta, S. Dev, \textit{Phys. Rev.}\textbf{D 
76}, 013006 (2007), hep-ph/0703005; M. Randhawa, G. Ahuja, M. Gupta, \textit{Phys. Lett.} \textbf{B 643}, 175-181 (2006), hep-ph/0607074.

\bibitem{8} E. I. Lashin and N. Chamoun, \textit{Phys. Rev.} \textbf{D 78}, 073002 (2008), hep-ph/0708.2423; E. I. Lashin, N.  Chamoun, \textit{Phys. Rev.} \textbf{D 80}, 093004 (2009), hep-ph/0909.2669; S. Dev, Surender Verma, Shivani Gupta and R. R. Gautam, \textit{Phys. Rev.} \textbf{D 
81}, 053010 (2010), hep-ph/1003.1006 
\bibitem{9} Biswajit Adhikari, Ambar Ghosal and Probir Roy, \textit{JHEP} \textbf{0910}, 040 (2009), hep-ph/0908.2686.
\bibitem{10} Wei Chao, Xiao-Gang He and Xue-Qian Li, \textit{Commun. Theor. Phys.} \textbf{45}, 1073-1084 (2006), hep-ph/0503285.
\bibitem{11} C. Amsler \textit{et al.}, [Particle Data Group], \textit{Phys. Lett.} \textbf{B 667}, 1 (2008) 
\bibitem{12} G. L. Fogli \textit{et al.}, \textit{Phys. Rev.}
\textbf{D 78}, 033010 (2008), hep-ph/0805.2517
\bibitem{13} E. Komatsu \textit{et al.}, \textit{Astrophys. J. Suppl.} \textbf{180}, 330-376 (2009), astro-ph/0803.0547  
\bibitem{14} H. V. Klapdor- Kleingrothaus, \textit{Nucl. Phys. Proc. Suppl.} \textbf{145}, 219 (2005).
\bibitem{15} Arnaboldi C \textit{et al.}, 2004a \textit{Nucl. Instrum. Meth.} \textbf{A 518}, 775.
\bibitem{16} Arnaboldi C \textit{et al.}, (CUORICINO collaboration) \textit{Phys. Lett.} \textbf{B 584}, 20 (2004).
\bibitem{17} I. Abt \textit{et al.}, (GERDA collaboration), hep-ex/ 0404039.
\bibitem{18} Sarazin X \textit{et al.}, 2000 Preprint hep- ex/ 0006031.
\bibitem{19} V. Lobashev  \textit{et al.}, \textit{Nucl. Phys.} \textbf{B} (Proc. Suppl.) \textbf{91}, 280 (2001); C. Weinheimer \textit{et al.}, \textit{Nucl. Phys. Proc. Suppl.} \textbf{118}, 279 (2003).    
\bibitem{20} P. Minkowski, \textit{Phys. Lett.}
\textbf{B 67}, 421 (1977); T. Yanagida, \textit{Proceedings of the
Workshop on the Unified Theory and the Baryon Number in the
Universe} (O. Sawada and A. Sugamoto, eds.), KEK, Tsukuba, Japan,
1979, p. 95: M. Gell-Mann, P. Ramond, and R. Slansky,
\textit{Complex spinors and unified theories in supergravity} (P.
Van Nieuwenhuizen and D. Z. Freedman, eds.), North Holland,
Amsterdam, 1979, p.315; R. N. Mohapatra and G. Senjanovic,
\textit{Phys. Rev. Lett.} \textbf{44}, 912 (1980).
\bibitem{21} Walter Grimus, Luis Lavoura, \textit{Phys. Lett.} \textbf{B 572}, 189-195 (2003), hep-ph/0305046.
\bibitem{22} Walter Grimus, Luis Lavoura, \textit{JHEP} \textbf{0508}, 013 (2005), hep-ph/0504153.
\bibitem{23} Walter Grimus, Luis Lavoura, \textit{Acta. Phys. Polon.} \textbf{B 32}, 3719 (2001), hep-ph/0110041.
\bibitem{24} Alexander Merle and Werner Rodejohann, \textit{Phys. Rev.} \textbf{D 73}, 073012 (2006), hep-ph/0603111; S. Dev and Sanjeev Kumar,
\textit{Mod. Phys. Lett.} \textbf{A 22}, 1401(2007),
hep-ph/0607048.
\bibitem{25} S. Dev, Sanjeev Kumar, Surender Verma and Shivani Gupta,
 \textit{Phys. Lett.} \textbf{B 656}, 79-82 (2007), hep-ph/0708.3321; Walter Grimus, Luis Lavoura, \textit{J. Phys.} \textbf{G 31}, 693-702 (2005) hep-ph/0412283; Sheldon Lee Glashow, hep-ph/0710.3719
\bibitem{26} R. N. Mohapatra and W. Rodejohann, \textit{Phys. Rev.} \textbf{D 72}, 053001 (2005), hep-ph/0507312.
\end{thebibliography}
\end{document}